\documentclass[10pt,twocolumn,letterpaper]{article}
\usepackage[pagenumbers]{include/cvpr/cvpr}

\usepackage{graphicx}
\usepackage{amsmath}
\usepackage{amssymb}
\usepackage{booktabs}
\usepackage{inconsolata}
\usepackage{multirow}
\usepackage{enumitem}
\usepackage{algorithm}
\usepackage{listings}
\usepackage{float}
\usepackage{xcolor}
\usepackage{colortbl}
\usepackage[pagebackref,breaklinks,colorlinks]{hyperref}
\usepackage{tikz}
\usepackage{indentfirst}

\def\paperID{27} 
\def\confName{CVMI}
\def\confYear{2024}

\newcommand{\xz}[1][scale=1]{
\begin{tikzpicture}[#1]
  \path[draw=black,fill=white,line cap=round,line join=round,line width=0.5pt] (0.2139, -0.0663) -- (0.1106, -0.0066) -- (0.1102, -0.1254) -- (0.2136, -0.1851) -- (0.2139, -0.0663);;
  \path[draw=black,fill=white,line cap=round,line join=round,line width=0.5pt] (0.1621, -0.0961) -- (0.0588, -0.0364) -- (0.0584, -0.1552) -- (0.1618, -0.2149) -- (0.1621, -0.0961);;
  \path[draw=black,fill=white,line cap=round,line join=round,line width=0.5pt] (0.1103, -0.1259) -- (0.007, -0.0662) -- (0.0066, -0.185) -- (0.11, -0.2447) -- (0.1103, -0.1259);;
\end{tikzpicture}
}
\newcommand{\xy}[1][scale=1]{
\begin{tikzpicture}[scale=1, transform shape]
  \path[draw=black,fill=white,line cap=round,line join=round,line width=0.5pt] (0.2129, -0.185) -- (0.1097, -0.1254) -- (0.007, -0.1851) -- (0.1102, -0.2447) -- (0.2129, -0.185);;
  \path[draw=black,fill=white,line cap=round,line join=round,line width=0.5pt] (0.2129, -0.1256) -- (0.1097, -0.066) -- (0.007, -0.1257) -- (0.1102, -0.1853) -- (0.2129, -0.1256);;
  \path[draw=black,fill=white,line cap=round,line join=round,line width=0.5pt] (0.2129, -0.0662) -- (0.1103, -0.0066) -- (0.0076, -0.0663) -- (0.1109, -0.1259) -- (0.2129, -0.0662);;
\end{tikzpicture}
}
\newcommand{\yz}[1][scale=1]{
\begin{tikzpicture}[scale=1, transform shape]
  \path[draw=black,fill=white,line cap=round,line join=round,line width=0.5pt] (0.1106, -0.0066) -- (0.0073, -0.0662) -- (0.0076, -0.185) -- (0.1109, -0.1254) -- (0.1106, -0.0066);;
  \path[draw=black,fill=white,line cap=round,line join=round,line width=0.5pt] (0.1617, -0.0365) -- (0.0585, -0.0961) -- (0.0588, -0.2149) -- (0.1621, -0.1553) -- (0.1617, -0.0365);;
  \path[draw=black,fill=white,line cap=round,line join=round,line width=0.5pt] (0.2129, -0.0663) -- (0.1097, -0.1259) -- (0.11, -0.2447) -- (0.2133, -0.1851) -- (0.2129, -0.0663);;
\end{tikzpicture}
}

\title{Super-resolution of biomedical volumes with 2D supervision}

\author{%
  Cheng Jiang\textsuperscript{1*}\quad
  Alexander Gedeon\textsuperscript{1*}\quad
  Yiwei Lyu\textsuperscript{1}\quad
  Eric Landgraf\textsuperscript{1}\quad
  Yufeng Zhang\textsuperscript{1}\\
  Xinhai Hou\textsuperscript{1}\quad
  Akhil Kondepudi\textsuperscript{1}\quad
  Asadur Chowdury\textsuperscript{1}\quad
  Honglak Lee\textsuperscript{1}\quad
  Todd Hollon\textsuperscript{1}\\[1ex]
  \textsuperscript{1}University of Michigan\quad
  \textsuperscript{*}Equal Contribution\\[1ex]
  \url{{chengjia, tocho}@umich.edu}\quad
  \url{https://mlins.org/msdsr/}
}

\begin{document}

\maketitle

\begin{abstract}
Volumetric biomedical microscopy has the potential to increase the diagnostic information extracted from clinical tissue specimens and improve the diagnostic accuracy of both human pathologists and computational pathology models. Unfortunately, barriers to integrating 3-dimensional (3D) volumetric microscopy into clinical medicine include long imaging times, poor depth/z-axis resolution, and an insufficient amount of high-quality volumetric data. Leveraging the abundance of high-resolution 2D microscopy data, we introduce masked slice diffusion for super-resolution (MSDSR), which exploits the inherent equivalence in the data-generating distribution across all spatial dimensions of biological specimens. This intrinsic characteristic allows for super-resolution models trained on high-resolution images from one plane (e.g., XY) to effectively generalize to others (XZ, YZ), overcoming the traditional dependency on orientation. We focus on the application of MSDSR to stimulated Raman histology (SRH), an optical imaging modality for biological specimen analysis and intraoperative diagnosis, characterized by its rapid acquisition of high-resolution 2D images but slow and costly optical z-sectioning. To evaluate MSDSR's efficacy, we introduce a new performance metric, SliceFID, and demonstrate MSDSR's superior performance over baseline models through extensive evaluations. Our findings reveal that MSDSR not only significantly enhances the quality and resolution of 3D volumetric data, but also addresses major obstacles hindering the broader application of 3D volumetric microscopy in clinical diagnostics and biomedical research.

\end{abstract}

\section{Introduction}
\begin{figure}[t!]
    \includegraphics[width=\columnwidth]{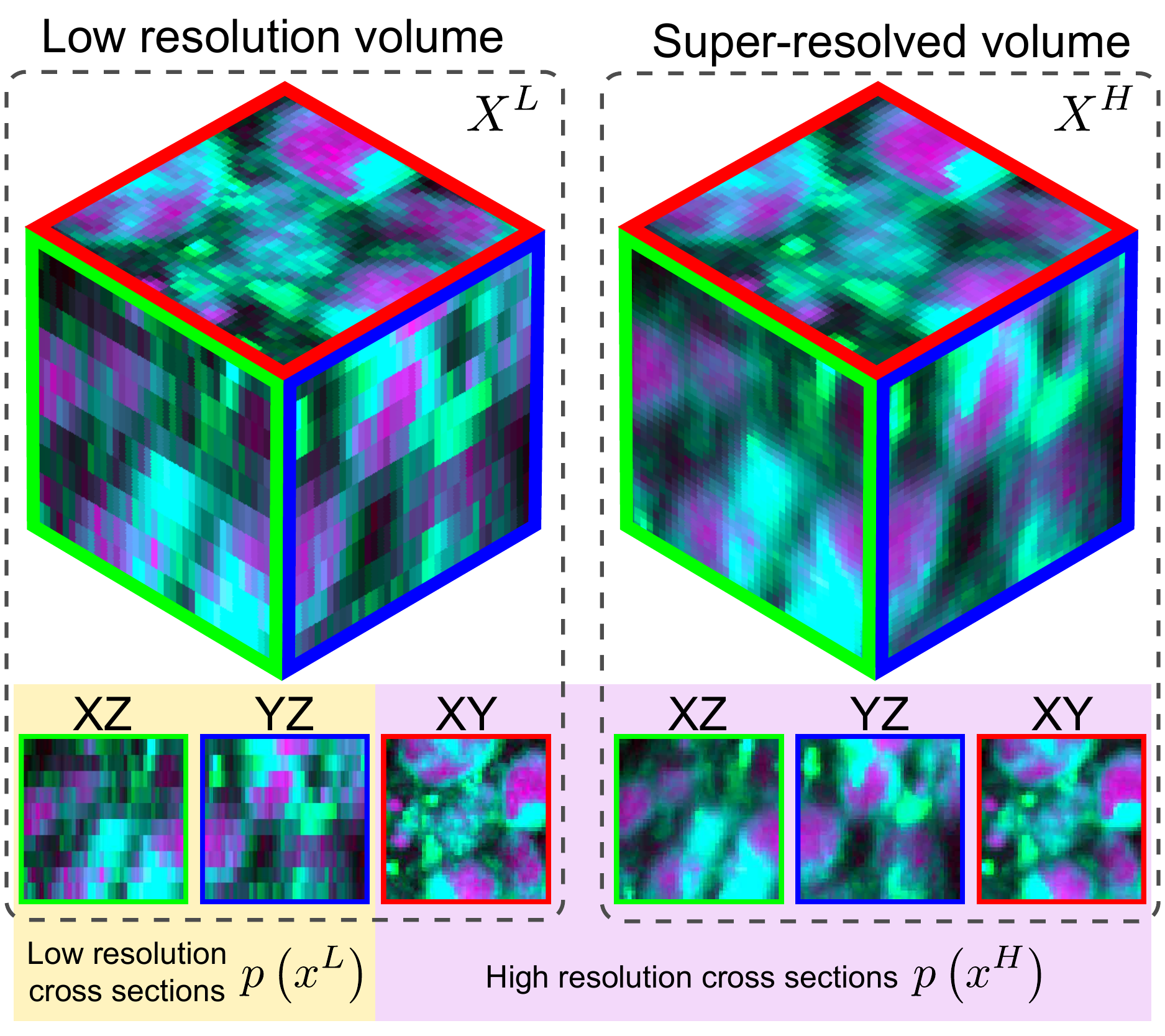}
    \caption{\textbf{Super-resolution of biomedical volumes with 2D supervision.} Volumetric microscopy images have a data distribution agnostic to tissue orientation and spatial dimension. Here, we present a method for leveraging this intrinsic characteristic by super-resolving low-resolution volumes using a conditional diffusion model trained on 2D high-resolution images.} 
    \label{fig:teaser}
\end{figure}

Biomedical microscopy is an essential imaging method and diagnostic modality in clinical medicine and biomedical research. Digital pathology and whole-slide images (WSIs) are now ubiquitous in computational pathology, leading to an increased role of computer vision and machine learning-based approaches for analyzing microscopy data. Recent research has shown that 3-dimensional (3D) volumetric microscopy can improve the diagnostic yield of surgical specimens and diagnostic accuracy of both human pathologists and computational pathology models \cite{glaser2017light, weigert2018content}. Diagnostic histoarchitectural and cytologic structures are three-dimensional, such as chromatin structure, microvilli, and perivascular rosette formation \cite{baloyannis2014fine}. The major barriers to integrating 3D volumetric microscopy into clinical medicine and biomedical research are (1) long imaging times, (2) poor depth (z-plane) resolution, and (3) insufficient high-quality 3D volumetric data. Importantly, high-quality, high-resolution, open-source 2D microscopy data is abundantly available, such as images from The Cancer Genome Atlas (TCGA), The Digital Brain Tumor Atlas (DBTA) \cite{roetzer2022digital}, and OpenSRH \cite{jiang2022opensrh}.

Here, we explore the open computer vision question of how to use high-resolution 2D microscopy data alone to improve the resolution of 3D volumetric microscopy, especially in the low-resolution z-plane or depth axis. We introduce \emph{masked slice diffusion for super-resolution (MSDSR)}, which leverages the observation that all three spatial dimensions share the same underlying data-generating distribution for biological specimens. For example, biological specimens sampled at the time of surgery for cancer diagnosis lack a spatial orientation and, therefore, microscopy images obtained in any imaging plane are valid images for cancer diagnoses. The lack of orientation (e.g. up-down, left-right, front-back) allows super-resolution models trained in any given 2D plane, such as XY, to generalize to any other, such as XZ or YZ.

We evaluate MSDSR using a label-free optical imaging modality that is used for biological specimen analysis and intraoperative diagnosis, called stimulated Raman histology (SRH) \cite{Freudiger2008-gj}. SRH is ideally suited for volumetric super-resolution because high-resolution 2D images are readily obtained, but z-sections through the depth of the specimen are slow and costly to obtain. Our major contributions are: 

\begin{enumerate}
    \item We introduce \emph{MSDSR}, a conditional diffusion-based, 3D volumetric super-resolution method that only requires 2D supervision.
    \item We introduce a new volumetric, unpaired, perceptual quality metric, \emph{SliceFID}.
    \item MSDSR outperforms both interpolation and UNet baselines on image quality metrics, including SliceFID.
\end{enumerate}

\begin{figure*}[htb!]
    \includegraphics[width=\textwidth]{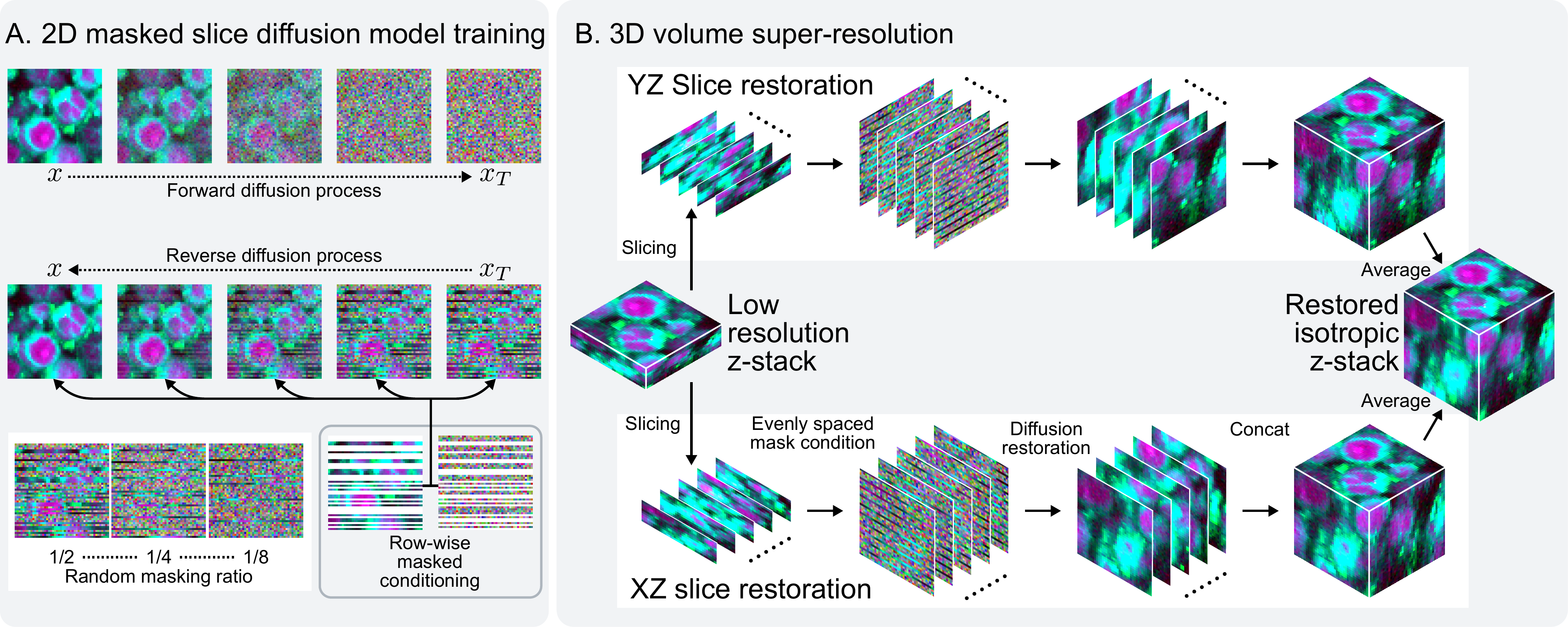}
    \caption{\textbf{MSDSR overview. A.} MSDSR is trained with a diffusion network conditioned on row-wise masks of the ground-truth high-resolution image. During the reverse diffusion process, a random masking ratio from 1/2 to 1/8 introduces these rows at random locations to give contextual structure when de-noising. The model then learns to interpolate the noised data in between the mask to produce a high-fidelity 2D image. \textbf{B.} During 3D inference, the low-resolution z-stack volume is sliced in both the XZ and YZ dimensions, producing low-resolution 2D images. The rows of these images are then treated as an evenly spaced mask interlacing random noise when individually fed into the model. These mixtures are then up-scaled by the model to produce high-resolution volumes and are then averaged together to generate a restored isotropic z-stack.}
    \label{fig:mech}
\end{figure*}

\section{Related Work}

\subsection{Denoising diffusion models} 

Generating high-fidelity, high-resolution images is a challenging task in computer vision. Generative models have recently gained popularity and media attention for natural image generation given a prompt or condition \cite{ramesh2022hierarchical,saharia2022photorealistic}. In particular, Denoising Diffusion Probabilistic Models (DDPMs) \cite{ho2020denoising} have shown state-of-the-art results on image synthesis \cite{dhariwal2021diffusion}, using a UNet architecture to iteratively transform random noise into the learned data distribution. However, these models suffer from heavy computational requirements compared to earlier methods such as variational autoencoders (VAEs) \cite{kingma2013auto} and generative adversarial networks (GANs) \cite{goodfellow2020generative} due to the iterative sampling process. Denoising diffusion implicit models (DDIMs) \cite{song2020denoising} accelerate the sampling process of DDPM using non-Markovian processes.  Latent diffusion models (LDMs) \cite{rombach2022high} further improve image quality and reduce computational requirements by performing the diffusion process on a latent space with lower dimensionality.

\subsection{Image super-resolution}

Super-resolution is the process of increasing the pixel resolution of an image. There exist several non-parametric methods for image super-resolution, such as nearest neighbor, linear, bilinear, and bicubic interpolation \cite{interpolationsurvey}. 

Regression-based methods such as SRCNN~\cite{dong2015image}, LIIF \cite{chen2021learning}, EDSR \cite{Lim_2017_CVPR_Workshops} and SwinIR \cite{liang2021swinir} directly learn mappings from low-resolution to high-resolution images with a pixel-wise loss. Generation-based methods are trained to generate a new image based on the input low-resolution image, such as SRGAN \cite{ledig2017photo}; CycleGAN \cite{zhu2017unpaired}, an image-to-image translation method, can be used to convert low-resolution images to high-resolution with unpaired data~\cite{kim2020unsupervised}. 

Image super-resolution with diffusion models is typically achieved with conditional diffusion, where the original image is used as a condition during the reverse diffusion process~\cite{saharia2021image}. Stable Diffusion \cite{rombach2022high} enabled efficient high-resolution image super-resolution by combining latent diffusion and conditional diffusion. Recent work combines conditional diffusion with GANs for better quality and faster inference speed~\cite{xiao2024single}. Aside from conditional diffusion, there are also diffusion-based super-resolution methods that either incorporate the input image in the denoising objective, such as DDRM~\cite{kawar2022denoising}, or directly perform iterative reverse diffusion on the low-resolution image, such as IDM~\cite{gao2023implicit}, and ResShift~\cite{yue2024resshift}.

\subsection{Super-resolution for biomedical imaging}
In radiological imaging, acquiring low-resolution images has the advantage of decreasing imaging time, radiation exposure, and motion artifacts. Super-resolution can then be used to interpolate information corrupted or lost during image acquisition. The majority of previous work on biomedical super-resolution has focused on radiological images, such as computed tomography (CT) or magnetic resonance imaging (MRI). Classical interpolation and reconstruction methods for medical images based on image processing methods and modeling the acquisition process have been studied \cite{greenspan2009super}. More recently, deep learning-based methods have gained prominence, including UNet \cite{ding2019deep,masutani2020deep}, autoencoder \cite{sander2022autoencoding,sander2021unsupervised}, and GAN frameworks \cite{mahapatra2017image,gu2020medsrgan,lyu2020mri,ahmad2022new,mahapatra2022mr,zhu2020csrgan}.

Progress has been made in enhancing 3D medical images, such as within radiology. Due to high computational complexity, many existing works combine 2D methods with some modifications for 3D consistency. Spatially aware interpolation network (SAINT) \cite{peng2020saint} utilized a 2D convolutional network for slice interpolation and a residual fusion network to ensure 3D consistency. Sood et al. \cite{sood20213d} used a GAN to generate novel fields-of-view using neighboring slices as conditioning, resulting in through-plane super-resolution and a more detailed 3D volume. Kudo et al. \cite{kudo2019virtual} applied a conditional GAN to enhance the generative diversity by incorporating the image information. Xia et al. \cite{xia2021super} combined optical flow interpolation with a GAN to generate an auxiliary image as supervision to guide image synthesis. Finally, ArSSR \cite{wu2022arbitrary} allowed arbitrary scale super-resolution of 3D MRIs via implicit neural representation.

Most recently, diffusion models have demonstrated remarkable effectiveness, and there have been a few studies to use diffusion models for 3D medical image super-resolution \cite{chung2022mr,chung2022score,chan2023super,wang2023inversesr,chung2023solving,lee2023improving,pan2023diffuseir}. \cite{chung2023solving,lee2023improving,pan2023diffuseir} are particularly relevant to our work, as they all attempted to use 2D diffusion models for 3D super-resolution. \cite{chung2023solving,lee2023improving} performed super-resolution on 3D MRI/CT by performing 2D super-resolution on perpendicular slices, but still require ground truth 3D high-resolution images to supervise model training. DiffuseIR \cite{pan2023diffuseir} trained a diffusion model for super-resolution using 2D slices, but did not perform 3D reconstruction of the entire volume. Furthermore, the previous works focused on single-channel MRI and electron microscopy data, whereas SRH generates multi-channel images.

\subsection{Deep Learning applications in SRH}

There have been many existing works on applying deep learning methods to SRH images, most of them focusing on classification tasks such as brain tumor subtype classification~\cite{Orringer2017-nn,hollon2020-ez}, molecular classification of diffuse glioma types~\cite{hollon2023artificial}, or whole-slide classification \cite{hou2024self, Jiang2023-rq}. Some prior works applied generative methods to denoising 2D SRH images~\cite{Manifold2019-ft, lyu2024step}. Lyu et al. \cite{lyu2024step} applied diffusion-based image restoration to 3D SRH z-stacks, but only to denoise each XY slice independently. To the best of our knowledge, this work is the first that attempts to super-resolve entire 3D SRH z-stack volumes.

\section{Methods}

The key motivation behind MSDSR is that different spatial dimensions of biomedical microscopy are equivalent and all views of 3D structures are orientation invariant. Thus, different slices of an isotropic 3D microscopy volume are 2D slices from the same underlying data-generating distribution. MSDSR leverages this observation and models the distribution using 2D images. The overall model architecture consists of the masked slice diffusion model and volume super-resolution inference, as described in figure \ref{fig:mech}.

Given a low-resolution volume $X^{L}\in\mathbb{R}^{ n \times n \times \ell}$, where $\ell<n$, we want to predict an isotropic high-resolution volume $X^{H}\in\mathbb{R}^{ n\times n\times n}$. Without high-resolution volume supervision, it is challenging to model  $p(X^{H})$ directly. We approximate the conditional probability of high-resolution volumes by modeling slices independently:
\begin{align}
    p\left(\left.X^{H}\right|X^{L}\right) &\approx \prod_{i=1}^{n} p\left(\left.X^{H}_{[i,:,:]}\right|X^{L}_{[i,:,:]}\right);\\
    p\left(\left.X^{H}\right|X^{L}\right) &\approx \prod_{j=1}^{n} p\left(\left.X^{H}_{[:,j,:]}\right|X^{L}_{[:,j,:]}\right),
\end{align}

\noindent where $X^{L}_{[i,:,:]}, X^{L}_{[:,j,:]} \in \mathbb{R}^{ \ell \times n}$ are YZ and XZ cross sections of the low-resolution volume $X^{L}$, respectively; and $X^{H}_{[i,:,:]}, X^{H}_{[:,j,:]} \in \mathbb{R}^{n\times n}$ are high-resolution YZ, XZ slices of $X^{H}$, respectively. Here, $p(X^{H}_{[i,:,:]}|X^{L}_{[i,:,:]})$ and $p(X^{H}_{[:,j,:]}|X^{L}_{[:,j,:]})$ are the conditional probability of the high-resolution YZ and XZ slices, given low-resolution YZ and XZ images with less rows in the Z dimension. Since XY, XZ, and YZ are from the same underlying distribution, it is equivalent to maximize the likelihood $p(x^H|x^L)$, where $x^H\in\mathbb{R}^{n\times n}$ is a high-resolution 2D image, and $x^L\in\mathbb{R}^{\ell\times n}$ is a lower resolution observation that can be simulated by downsampling or masking at training time. Since high-resolution 3D data is challenging and expensive to acquire, training with 2D images allows us to learn a better model by leveraging more data.

\subsection{Masked slice diffusion} \label{sec:methods.msd}
We train a DDPM to generate a high-resolution 2D image $x^H$, conditioned on the paired low-resolution image $x^L$. During training, we simulate the paired low-resolution image by removing rows of the high-resolution image. We obtain high-resolution images from the XY plane, and the trained model still applies well to XZ and YZ super-resolution due to the dimensional equivalence of SRH microscopy. Following the key results in \cite{ho2020denoising}, forward diffusion is a fixed process that gradually adds Gaussian noise to the image following a noise schedule $\beta$, for a total of $T$ steps. At each step $t$, 
\begin{equation}
    x^H_t\sim\mathcal{N}\left(\sqrt{\bar\alpha_t}x^H,\left(1-\bar\alpha_t\right)\boldsymbol{I}\right),
\end{equation}
\noindent where $\alpha_t=1-\beta_t$, and $\bar\alpha_t=\prod_{s=1}^{t}\alpha_s$.

During the reverse diffusion process, we condition $x^H_t$ by interlacing it with the simulated low-resolution image. We sample $\ell\sim\text{Uniform}([\ell_\text{min}, \ell_\text{max}])$ number of rows to include as the condition, where $\ell_\text{min}$,$\ell_\text{max}$ are hyperparameters such that $0<\ell_\text{min}<\ell_\text{max}<n$. $S\sim\text{Uniform}([1, n], \ell)$ is a set of random $\ell$ indices drawn without replacement for each row to be interlaced into $x^H_t$. We create a row-wise binary mask $b$ to combine the partially denoised image at timestep $t$ and the low-resolution image condition:
\begin{align}
    b &= \left[\mathbf{1}_{S}(1), \dots, \mathbf{1}_{S}(n)\right]^\top \\
    c(x_t^H,x^H,b) &= b * x^H + (1-b) * x_t^H,
\end{align}
where $\mathbf{1}(\cdot)$ is the indicator function, and $*$ denotes element-wise multiplication with broadcasting.

The masked slice diffusion model $\epsilon_\theta$ in the reverse diffusion process is optimized using the variational lower bound on the negative log-likelihood with the objective function (i.e. the simplified objective from~\cite{ho2020denoising}): 
\begin{equation}\label{eq:loss}
\mathcal{L}=\mathbb{E}_{x^H, b, \epsilon \sim \mathcal{N}(\mathbf{0}, \mathbf{I}), t}\left[\left\| b*\epsilon -\epsilon_\theta\left(c(x_t^H,x^H, b), t\right)\right\|\right].
\end{equation}

\subsection{Volume super-resolution inference}

To generate high-resolution volumes, we use our masked slice diffusion model to infer high-quality YZ and XZ slices along the X and Y axes:
\begin{align}
    \hat{X}^{H}_{YZ} = \text{Concat}\left[f\left(X_{[1,:,:]}^{L}\right), \dots, f\left(X_{[n,:,:]}^{L}\right) \right];\\
    \hat{X}^{H}_{XZ} = \text{Concat}\left[f\left(X_{[:,1,:]}^{L}\right), \dots, f\left(X_{[:,n,:]}^{L}\right) \right],
\end{align}
\noindent where $f$ is the full reverse diffusion restoration process of our masked slice diffusion model, including sampling $x_T$, denoising and interlacing the observed low-resolution image at each time step. During the restoration process, each of the $m$ slices along an axis is super-resolved independently. This independence does not reflect the physical structure we are trying to render, as neighboring slices are correlated. As a result, concatenation artifacts may form along the slices orthogonal to the inference axes. To eliminate the independence of inferences between planes, we use averaging to combine the volumes super-resolved in both directions:
\begin{equation}
    \hat{X}^{H} = \frac{\hat{X}^{H}_{YZ} + \hat{X}^{H}_{XZ}}{2}.
\end{equation}
As shown in section \ref{sec:results}, this straightforward method achieves good empirical results in reducing inconsistencies and concatenation artifacts on the super-resolved volumes. A pseudocode of the MSDSR inference process is in algorithm \ref{alg:inf}.

\begin{algorithm}[h!]
\caption{MSDSR volume inference in PyTorch style.}
\label{alg:inf}
\definecolor{codeblue}{rgb}{0.0,0.5,0.0}
\definecolor{codekw}{rgb}{0.85, 0.18, 0.50}
\lstset{
  basicstyle=\fontsize{8pt}{8pt}\ttfamily\selectfont,
  columns=fullflexible,
  breaklines=true,
  captionpos=b,
  commentstyle=\color{codeblue},
  keywordstyle=\color{codekw},
}

\begin{lstlisting}[language=python,showstringspaces=false]
def superresolve_along_axis(x):
    # x: transposed low res image of shape [n 3 l n]
    
    high_res_ims = []
    mask = arange(0, n, n // l) # assume l is a factor of n
    
    for i in range(n): # for each low res slice

        # interlace random noise with the observation
        x_T = randn_like(x[i])
        x_T[:, mask, :] = x[i]

        # full reverse diffusion restoration process
        high_res_ims.append(msdsr.restore(x_T))

    return stack(high_res_ims)

def superresolve_volume(xl):
    # xl: low res image of shape [3 n n l] (CHWZ)

    # transpose and superresolve xl in XZ slices
    xl_xz = rearrange(xl, "c h w z -> h c z w")
    xh_xz = superresolve_along_axis(xl_xz)
    xh_xz = rearrange(xh_xz, "h c z w -> c h w z")

    # transpose and superresolve xl in YZ slices
    xl_yz = rearrange(xl, "c h w z -> w c z h")
    xh_yz = superresolve_along_axis(xl_yz)
    xh_yz = rearrange(xh_yz, "w c z h -> c h w z")

    # return high resolution volume
    return (xh_yz + xh_xz) / 2
\end{lstlisting}
\end{algorithm}

\section{Experimentation}

We evaluated MSDSR on a z-stacked volumetric stimulated Raman histology (SRH) dataset and compared the results to interpolation and UNet baselines.

\subsection{Data description} \label{sec:datadesc}
Our z-stacked SRH dataset was collected using tumor specimens from patients who underwent brain tumor biopsy or resection at the University of Michigan. This study was approved by the Institutional Review Board (HUM00083059), and informed consent was obtained from each patient before imaging. The z-stacked SRH imaging of fresh surgical specimens follows the imaging protocol described in \cite{jiang2022opensrh}. Each slide has a $0.5 \times 0.5\ \mathrm{mm}^2$ field of view and a $1000 \times 1000$ pixel resolution. The slides are imaged at an initial depth of $20\ \mathrm{\mu m}$, and the laser focus is adjusted for each subsequent slice for virtual sectioning. Each z-stacked volume has a z-resolution of $1\ \mathrm{\mu m}$ for 20 z-sections. These z-stacks are not isotropic due to the physical limitations of optical sectioning, imaging time, and cost. Our data ingest and image processing pipeline also follows \cite{jiang2022opensrh}, with each whole-slide image volume being patched into $256\times 256\times 20$ pixels$^3$ tiles. Every patch is also subsequently denoised using RSCD \cite{lyu2024step} before model development and validation.

Our z-stacked dataset consists of 1129 whole-slide images from a total of 300 patients, spanning a wide range of brain tumor types, including glioma, meningioma, metastases, pituitary adenoma, schwannoma, and other less common central nervous system tumors. The dataset is split into training and validation sets, with 241 patients for model training and 59 patients for validation. Additionally, a key advantage of MSDSR is that it does not require training data to be volumetric. As a result, we also utilized a larger 2D SRH dataset consisting of SRH images from 1021 patients for masked slice training.

\subsection{Implementation details}
\paragraph{MSDSR architecture and training.} Our MSDSR model is implemented using DDPM with 274M parameters. Based on the depth field of view of our SRH images, we crop our input images to $48\times 48$. Our DDPM model utilizes a cosine noise scheduler with $T=1000$ steps. The model is supervised with an L1 loss and optimized using AdamW, with a base learning rate of $10^{-7}$ and a cosine learn rate scheduler with a 10\% warmup. The model was trained until convergence with an effective batch size of 256.

At training time, we condition the reverse diffusion process by randomly masking rows of high-resolution images, as described in section \ref{sec:methods.msd}. The number of rows to be used as the condition is randomly drawn from a uniform distribution, with $\ell_\text{min}=5$ and $\ell_\text{max}=20$. These parameters were selected based on our z-stacked SRH dataset: $\ell_\text{max}=20$ matches the resolution of existing data (at half resolution relative to XY slices); and $\ell_\text{min}=5$ matches data collected with 4$\times$ speed up, resulting in a z-resolution that is 1/8 of X- and Y-resolution.

\paragraph{MS-UNet baseline.} We applied our masked slice training approach to a UNet architecture (MS-UNet). We use the same strategy to train the model using high-resolution images interlaced with random noise rows as input. An L1 loss between the prediction and the ground truth high-resolution 2D image was used to supervise model training. The UNet model was trained with a base learning rate of $10^{-3}$, with other hyperparameters kept the same as MSDSR training.

\paragraph{End-to-end (E2E) UNet baseline.} In addition to masked slice training, we also trained an end-to-end UNet to interpolate two different slides. The end-to-end UNet has the same architecture as the MS-UNet, except it takes a six-channel input consisting of two RGB slices $X_i$ and $X_{i+2}$. The model outputs the slice $X_{i+1}$ between the two input slices and is supervised with an L1 loss function. All hyperparameters are the same as MS-UNet training. 

\begin{table*}[!h] \centering
    \begin{tabular}{c|cc|cc|cc}  \toprule
    & \multicolumn{2}{c|}{2$\times$ super-resolution} & \multicolumn{2}{c|}{4$\times$ super-resolution} & \multicolumn{2}{c}{8$\times$ super-resolution} \\
    & FID  & SSIM & FID & SSIM & FID & SSIM\\
    \midrule
    NN              &         90.0  &         0.714  &         244.8 &         0.426  &         450.4 &         0.228\\
    Bilinear        &         31.6  &         0.727  &         134.0 &         0.442  &         314.6 &         0.241\\
    MS-UNet         &         24.9  & \textbf{0.825} &         66.8  & \textbf{0.628} &         163.4 & \textbf{0.419}\\
    MSDSR (Ours)    & \textbf{21.0} &         0.678  & \textbf{21.5} &         0.486  & \textbf{22.3} &         0.284\\
    \bottomrule
    \end{tabular}
\caption{\label{tab:metrics_2dpair}\textbf{Paired 2D evaluation metrics}. We present the FID and SSIM scores of our models and baselines on 2D paired data with a scaling factor ranging from 2$\times$ to 8$\times$. These comparisons come from inference on high-resolution 2D images. While MS-UNet achieves a higher SSIM, images super-resolved by MS-UNet are perceptually blurry. NN, nearest neighbor, bilinear, bilinear interpolation.}
\end{table*} 

\subsection{Evaluation protocol}
\paragraph{Paired 2D evaluation.} We evaluate the model by comparing super-resolved images with their high-resolution ground-truths. We uniformly mask 24, 36, and 42 rows from high-resolution 2D images for 2$\times$, 4$\times$, and 8$\times$ super-resolution tasks, respectively. We report FID and SSIM as quantitative metrics.

\paragraph{Unpaired 3D evaluation.} We evaluate 3D super-resolution using the z-stacked SRH images as described in section \ref{sec:datadesc}. It is challenging to apply FID, an unpaired metric, to volumetric SRH data because it requires a learned embedding space to measure the similarity of the data distribution. Prior work, such as 3D-FID \cite{dorjsembe2024conditional} and FVD \cite{unterthiner2019fvd}, use specialized volumetric feature extractors to compute embeddings, which is not feasible for our z-stacked SRH dataset. Thus, we propose an evaluation metric, \emph{SliceFID}, to measure the quality of restored z-stack microscopy data.

SliceFID is motivated by the domain knowledge that XY, YZ, and XZ slices of an isotropic 3D volume are different views of the same underlying biological structures and follow the same data distribution. Therefore, we can utilize the FID to compute the distance between a set of high-quality two-dimensional images and each of the XY, YZ, and XZ slices of the generated image. The FID score in each axis informs the perceptual quality of images along each axis, and SliceFID is defined as the average of these per-axis metrics:
\begin{multline}
    \text{SliceFID}(x^H,\hat{X}) = \frac{1}{3}\left[\text{FID}(x^H,\text{slice}_\text{XY}(\hat{X})) + \right.\\ \left.\text{FID}(x^H,\text{slice}_\text{XZ}(\hat{X})) + \text{FID}(x^H,\text{slice}_\text{YZ}(\hat{X}))\right],
\end{multline}
where $x^H\in\mathbb{R}^{k\times h\times w}$ is a set of $k$ high-quality ground truth 2D images, $\hat{X}\in\mathbb{R}^{m\times h\times w\times z}$ is a set of $m$ super-resolved volumes, and $\text{slice}_\text{XY}(\hat{X})\in\mathbb{R}^{mz\times h\times w}$, $\text{slice}_\text{XZ}(\hat{X})\in\mathbb{R}^{mh\times w\times z}$ and $\text{slice}_\text{YZ}(\hat{X})\in\mathbb{R}^{mw\times h\times z}$ are the generated volumes in XY, XZ, and YZ slices, respectively.

\section{Results} \label{sec:results}

\subsection{MSDSR paired 2D evaluation}
\begin{figure}[tb!]
    \includegraphics[width=\columnwidth]{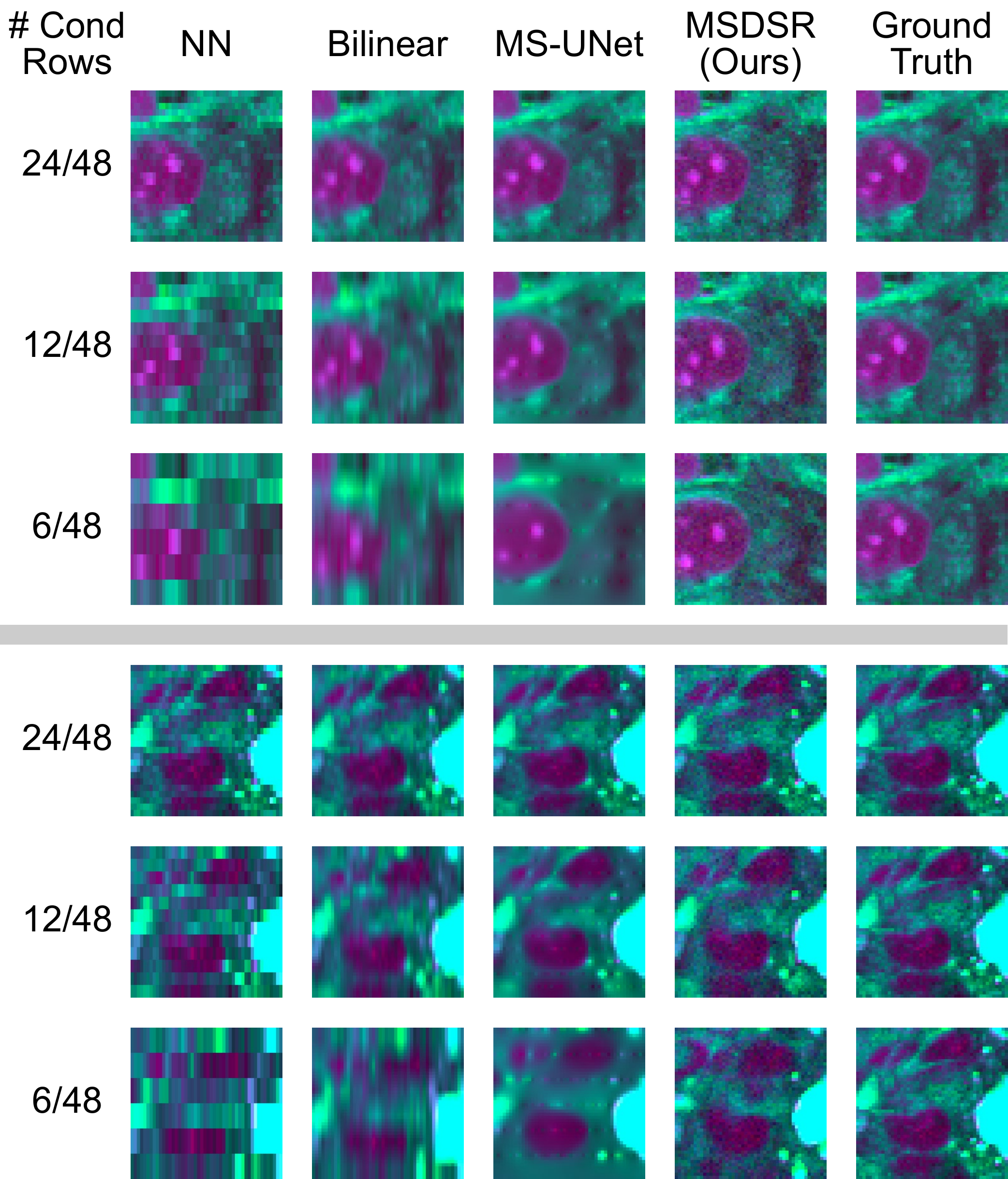}
    \caption{\textbf{Paired 2D evaluation.} We compare the images generated by MSDSR and other baselines to the paired ground truth image. \# cond rows, number of conditioning rows, NN, nearest neighbor, bilinear, bilinear interpolation.} 
    \label{fig:fig.2d_eval}
\end{figure}

In this section, we evaluate MSDSR on a paired 2D super-resolution task and compare it to interpolation and UNet baselines. Table \ref{tab:metrics_2dpair} summarizes the quantitative metrics, and a panel of super-resolved examples with various numbers of conditioning rows is shown in figure \ref{fig:fig.2d_eval}.

Quantitatively, MSDSR outperforms all baseline methods on FID across all super-resolution tasks. MS-UNet achieves the best SSIM metric across all tasks but produces perceptually blurry images. MSDSR outperforms nearest neighbor and bilinear interpolation baselines in SSIM with a larger super-resolution scaling factor (4$\times$ and 8$\times$).

Visually, MSDSR generates high-fidelity images similar to the paired ground truth, where NN and bilinear interpolation generate images with significant blurring and artifacts. MS-UNet produces overly smooth images, with missing details in cellular structures and background objects. MSDSR consistently recovers relevant cellular features (e.g., shape, chromatin, cytoplasm) with increasingly lower-resolution input images while maintaining realistic details, making it the only robust super-resolution method benchmarked. 

\begin{table*}[!h] \setlength{\tabcolsep}{4pt}
\centering\begin{tabular}{c|cccc|cccc|cccc}  \toprule
& \multicolumn{4}{c|}{2$\times$ super-resolution FID} & \multicolumn{4}{c|}{4x super-resolution FID} & \multicolumn{4}{c}{8$\times$ super-resolution FID} \\
& XY\xy & YZ\yz & XZ\xz & SliceFID & XY\xy & YZ\yz & XZ\xz & SliceFID & XY\xy & YZ\yz & XZ\xz &  SliceFID \\
\midrule
NN           &         24.7  &         197.9 &         166.2 &         129.6 & \textbf{24.8} &         357.7 &         354.6 &         245.7 & \textbf{25.1} &         530.8 &         447.2 &         334.4\\
Bilinear     &         30.6  &         62.0  &         60.6  &         51.1  &         29.3  &         179.1 &         162.7 &         123.7 &         27.3  &         355.0 &         357.3 &         246.5\\
E2E UNet     &         43.9  &         73.0  &         63.4  &         60.1  &         58.7  &         179.8 &         167.9 &         135.4 &         70.9  &         311.9 &         298.7 &         227.2\\
MS-UNet      &         38.8  &         46.2  &         55.0  &         46.6  &         58.0  &         98.9  &         95.8  &         84.2  &         103.9 &         224.4 &         189.6 &         172.7\\
MSDSR (Ours) & \textbf{16.2} & \textbf{28.9} & \textbf{31.4} & \textbf{25.5} &         25.5  & \textbf{37.1} & \textbf{35.8} & \textbf{32.8} &         107.4 & \textbf{61.9} & \textbf{56.7} & \textbf{75.3}\\
\bottomrule
\end{tabular}
\caption{\label{tab:3d_metrics}\textbf{3D super-resolution metrics.} We present the SliceFID score and its components for MSDSR and MS-UNet along with our baseline methods. NN, nearest neighbor, bilinear, bilinear interpolation, E2E UNet, end-to-end UNet.}
\end{table*} 

\begin{figure}[t!]
    \includegraphics[width=\columnwidth]{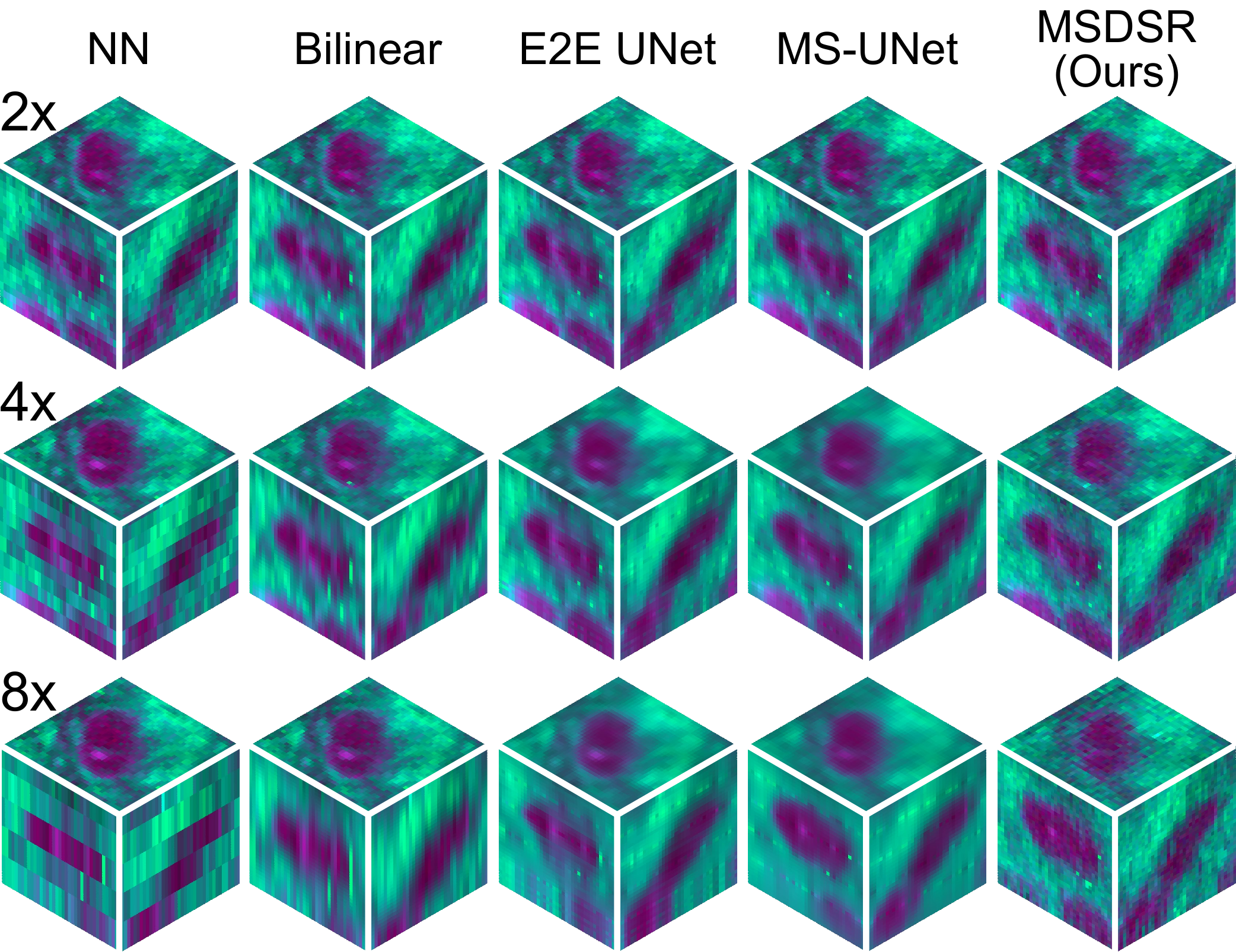}
    \caption{\textbf{3D super-resolution results.} We compare 3D volumetric super-resolution inference across three different input scalings. NN, nearest neighbor, bilinear, bilinear interpolation, E2E UNet, end-to-end UNet.} 
    \label{fig:3d_eval}
\end{figure}
\begin{figure}[t!]
    \includegraphics[width=\columnwidth]{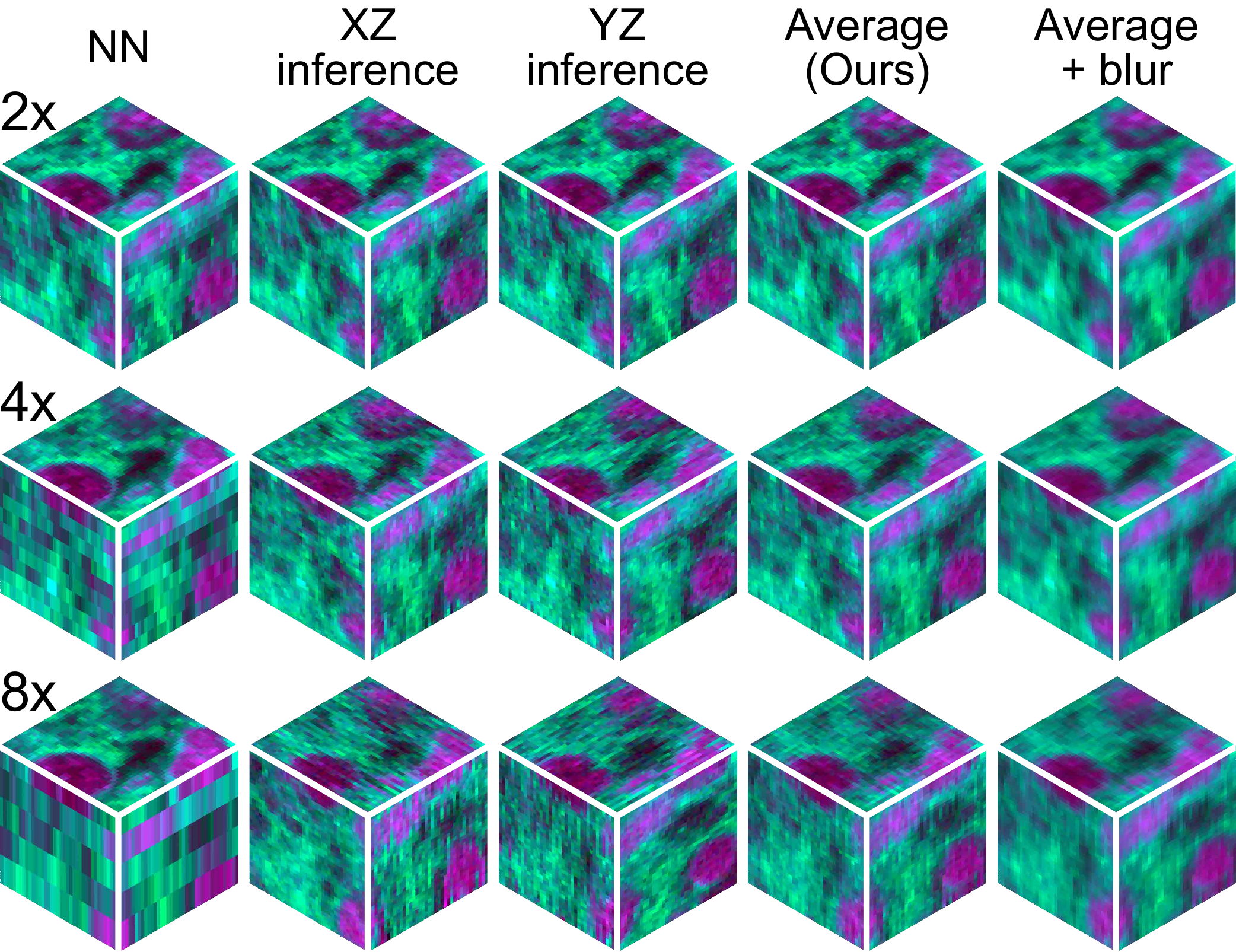}
    \caption{\textbf{Ablation study on inference direction and Gaussian blur.} We compare 3D volumetric super-resolution inference using ablated models across three different input scalings. NN, nearest neighbor, average, averaging XZ and YZ inference.} 
    \label{fig:3d_ablations}
\end{figure}

\subsection{MSDSR 3D Inference Evaluation}

To evaluate the quality of generated z-stack volumes, we use SliceFID to assess the volumetric image quality along each of the XY, YZ, and XZ planes, as well as their average for a holistic evaluation. Table \ref{tab:3d_metrics} shows SliceFID and its components, and figure \ref{fig:3d_eval} shows a sample super-resolved volume for each model, across three different super-resolution scaling factors.

Overall, MSDSR achieves the best SliceFID score across all super-resolution tasks. Non-parametric interpolation methods (i.e., NN and bilinear) have a low FID score along the XY plane because they naively interpolate information between the XY slices in the z-direction, leaving the input data intact. As a result, YZ and XZ images generated by these methods are unrealistic and contain jittering and stretching artifacts due to the sparsity of the input data on the plane. Both UNet-based models achieved better performance compared to the non-parametric methods, but they both generate overly smooth images, lacking detail in the nuclei and cytoplasm of the cells. The naive E2E UNet had worse results because it only models images in the XY plane and is only trained using a fixed 2-micron interval. When inferencing with a lower resolution condition, the E2E model relies on a recursive strategy to predict intermediate slices and amplifies any mistakes made during the process. While MS-UNet performs better, it still generates smoothed images that are not realistic. In comparison, MSDSR performs the best and generates high-fidelity volumes, with cellular and background textures remaining available. As condition image resolution decreases, MSDSR still maintains consistent and reasonable images, albeit at a mildly worse quality.

\begin{table*}[!t] \setlength{\tabcolsep}{4pt}
\centering\begin{tabular}{c|cccc|cccc|cccc}  \toprule
& \multicolumn{4}{c|}{2$\times$ super-resolution FID} & \multicolumn{4}{c|}{4$\times$ super-resolution FID} & \multicolumn{4}{c}{8$\times$ super-resolution FID} \\
& XY\xy & YZ\yz & XZ\xz & SliceFID & XY\xy & YZ\yz & XZ\xz & SliceFID & XY\xy & YZ\yz & XZ\xz &  SliceFID \\\midrule
MSDSR-XZ \xz    & 14.5 &         28.1  & \textbf{27.9} & 23.5 &  23.5 &         38.8  & \textbf{28.1}& 30.1 & 88.5  &        103.9  & \textbf{25.6} & 72.7\\
MSDSR-YZ \yz    & 15.1 & \textbf{27.1} &         28.0  & 23.4 &  26.6 & \textbf{28.6} &         36.3 & 30.5 & 102.8 & \textbf{27.3} &         100.7 & 76.9\\\bottomrule
\end{tabular}
\caption{\textbf{3D ablation FID metrics.} We present the SliceFID score and its components for MSDSR ablated for inference in only a single dimension. MSDSR-XZ represents performing inference solely in the XZ dimension, with the analogous MSDSR-YZ. We can observe a significant performance drop on the planes orthogonal to the inference plane, most likely due to a stitching artifact.}\label{tab:3dablation_metrics}
\end{table*}

\definecolor{greendelta}{HTML}{5C940D}
\definecolor{reddelta}{HTML}{C92A2A}
\newcommand{\arrangered}[2]{#1 {\color{reddelta} #2}}
\newcommand{\arrangegreen}[2]{#1 {\color{greendelta} #2}}
\newcommand{\blurdelta}{($\Delta$)}
\begin{table*}[!t]
\setlength{\tabcolsep}{4pt}
\centering\begin{tabular}{cc|cccc}  \toprule
SR scale & Method & XY\xy \blurdelta & YZ\yz \blurdelta & XZ\xz \blurdelta & SliceFID \blurdelta\\\midrule
   & MSDSR-XZ\xz + blur & \arrangered{23.5}{(+9.0)}  & \arrangered{34.3}{(+6.2)}  & \arrangered{37.9}{(+10.0)} & \arrangered{31.9}{(+8.4)}  \\
2$\times$  & MSDSR-YZ\yz + blur & \arrangered{24.0}{(+9.0)}  & \arrangered{38.5}{(+11.4)} & \arrangered{37.3}{(+9.4)}  & \arrangered{33.3}{(+9.9)}  \\
   & MSDSR + blur       & \arrangered{30.7}{(+14.5)} & \arrangered{44.2}{(+15.3)} & \arrangered{45.4}{(+14.0)} & \arrangered{40.1}{(+14.6)} \\\midrule
   & MSDSR-XZ\xz + blur & \arrangered{26.3}{(+2.8)} & \arrangered{39.1}{(+0.3)} & \arrangered{34.7}{(+6.7)} & \arrangered{33.4}{(+3.3)}\\
4$\times$  & MSDSR-YZ\yz + blur & \arrangered{28.5}{(+1.9)} & \arrangered{36.1}{(+7.5)} & \arrangered{41.0}{(+4.7)} & \arrangered{35.2}{(+4.7)}\\
   & MSDSR + blur       & \arrangered{32.7}{(+7.2)} & \arrangered{43.5}{(+6.3)} & \arrangered{45.0}{(+9.2)} & \arrangered{40.4}{(+7.6)}\\\midrule
   & MSDSR-XZ\xz + blur & \arrangegreen{72.5 }{(-16.0)} & \arrangegreen{88.5}{(-15.3)} & \arrangered{30.2}{(+4.6)}     & \arrangegreen{63.7}{(-8.9)}\\
8$\times$ & MSDSR-YZ\yz + blur & \arrangegreen{84.7}{(-18.1)}  & \arrangered{30.9}{(+3.5)}    & \arrangegreen{79.7}{(-21.0)}  & \arrangegreen{65.1}{(-11.9)}\\
   & MSDSR + blur       & \arrangegreen{101.2}{(-6.2)}  & \arrangered{62.1}{(+0.2)}    & \arrangered{59.5}{(+2.8)}     & \arrangegreen{74.3}{(-1.1)}\\
\bottomrule
\end{tabular}

\caption{\textbf{3D Gaussian blur ablation FID metrics.} $3\times 3\times 3$ Gaussian blur post-processing degrades the model performance for 2$\times$ and 4$\times$ super-resolution, but offers a boost for the 8$\times$ task, especially in the planes orthogonal to the inference plane. Differences to FID scores before Gaussian blurring are reported in parentheses ($\Delta$). SR scale, super-resolution scale.} \label{tab:3dablation_metrics_blur}
\end{table*}

\paragraph{Ablation studies.}
We investigate the effect of averaging inferences from both XZ and YZ directions, as well as using Gaussian blur as a post-processing step. Quantitative SliceFID metrics are reported in tables \ref{tab:3dablation_metrics} and \ref{tab:3dablation_metrics_blur}, respectively. Examples of super-resolved volumes are shown in figure \ref{fig:3d_ablations}. Inferencing along a single direction results in noticeable stitching artifacts for 4$\times$ and 8$\times$ super-resolved images, especially in the planes orthogonal to the inference plane, both visually and quantitatively. This results from aggregating independent inferences that are plausible solutions during individual slice inference, given a lower resolution condition. Averaging XZ and YZ alleviates the artifact, but does not completely remove it. Applying Gaussian blur is another way to reduce the stitching artifact, especially in 8$\times$ super-resolution task, but it also reduces the overall sharpness of the prediction.

\section{Conclusion}

Our study presents a novel approach, masked slice diffusion for super-resolution (MSDSR), to enhance the resolution of 3D volumetric biomedical images utilizing 2D supervision. We demonstrated that MSDSR can leverage the inherent similarity in data-generating distributions across spatial dimensions of biological specimens, enabling effective generalization from 2D to 3D. Our proposed method significantly surpasses traditional interpolation and UNet baselines across various image quality metrics, notably through our newly introduced SliceFID metric, emphasizing MSDSR's efficacy in generating high-quality, realistic volumetric reconstructions from low-resolution inputs.

\paragraph{Limitations.} While MSDSR has shown promising results, it is not without limitations. The primary challenge lies in the method's current reliance on synthesizing slices independently, which can lead to inconsistencies in 3D volumetric reconstructions. This approach, while effective for enhancing individual slices, does not fully exploit the spatial correlations inherent in 3D structures, potentially affecting the consistency of the reconstructed volumes. Furthermore, the computational demands of our method, particularly for high-resolution volumetric data, pose challenges for real-time clinical applications.

\paragraph{Broader impact.} The broader impact of MSDSR extends beyond the technical achievements in biomedical imaging. By significantly improving the resolution and quality of 3D volumetric microscopy, our work has the potential to advance diagnostic accuracy, enhance the understanding of complex biological structures, and facilitate the development of novel therapeutic strategies. Furthermore, by reducing the dependency on high-resolution 3D data, MSDSR can democratize access to advanced imaging technologies, particularly in resource-constrained settings, ultimately contributing to the global efforts to bridge the healthcare divide.

\section*{Acknowledgements and Competing Interests}
We would like to thank Karen Eddy, Lin Wang, and Hubert Zhang for their administrative support and data collection efforts.

This work was supported, in part, by the National Institutes of Health (NIH) grants F31NS135973 (C.J.), T32GM141746 (C.J.), and K12NS080223 (T.H.). This work was also supported, in part, by the Chan Zuckerberg Foundation (CZI) Advancing Imaging Through Collaborative Project grant (T.H.), the Cook Family Brain Tumor Research Fund (T.H.), the Mark Trauner Brain Research Fund (T.H.), the Zenkel Family Foundation (T.H.), Ian’s Friends Foundation (T.H.) and the UM Precision Health Investigators Awards grant program (T.H.).

T.H. is a shareholder of Invenio Imaging, Inc., a company developing SRH microscopy systems.

\bibliographystyle{include/cvpr/ieeenat_fullname}
\bibliography{latex/paperpile,latex/additional}
\end{document}